\documentclass[doublecol]{epl2} 
\usepackage{amssymb}
\usepackage{color}

\title{Model Selection based on the Angular-Diameter Distance \\ to the Compact Structure in Radio Quasars}
\shorttitle{Angular-Diameter Distance} 

\author{F. Melia\inst{1}}
\shortauthor{F. Melia}

\institute{                    
  \inst{1} Department of Physics, The Applied Math Program, and Department of Astronomy, \\
The University of Arizona, AZ 85721, USA}
\pacs{95.36.+x}{Dark Energy}
\pacs{98.80.-k}{Cosmology}
\pacs{98.80.Jk}{Relativistic Astrophysics}

\abstract{Of all the distance and temporal measures in cosmology, the angular-diameter
distance, $d_A(z)$, uniquely reaches a maximum value at some finite redshift
$z_{\rm max}$ and then decreases to zero towards the big bang. This effect has
been difficult to observe due to a lack of reliable, standard rulers, though
refinements to the identification of the compact structure in radio 
quasars may have overcome
this deficiency. In this {\it Letter}, we assemble a catalog of 140 such
sources with $0\lesssim z\lesssim 3$ for model selection and the measurement
of $z_{\rm max}$. In flat $\Lambda$CDM, we find that $\Omega_{\rm m}=
0.24^{+0.1}_{-0.09}$, fully consistent with {\it Planck}, with $z_{\rm max}=1.69$.
Both of these values are associated with a $d_A(z)$ indistinguishable
from that predicted by the zero active mass condition, $\rho+3p=0$, in terms
of the total pressure $p$ and total energy density $\rho$ of the cosmic
fluid. An expansion driven by this constraint, known as the $R_{\rm h}=ct$
universe, has $z_{\rm max}=1.718$, which differs from the measured value
by less than $\sim 1.6\%$. Indeed, the Bayes Information Criterion favours
$R_{\rm h}=ct$ over flat $\Lambda$CDM with a likelihood of $\sim 81\%$
versus $19\%$, suggesting that the optimized parameters in {\it Planck}
$\Lambda$CDM mimic the constraint $p=-\rho/3$.}

\begin{document}

\maketitle

\section{Introduction}
The luminosity distance is used often in cosmology for measurements involving 
standard candles, such as Type Ia SNe \cite{Kowalski:2008} and gamma ray bursts 
\cite{Wei:2013}. By comparison, the angular-diameter distance, $d_A(z)$, applicable 
to objects whose diameter (preferably a `standard ruler') is known, is used only
sparingly, given the relative paucity of such sources and complications arising 
from size evolution with redshift. More typically, sources used to
measure $d_A(z)$ have been restricted to narrow ranges in redshift, mitigating
their possible impact on revealing the geometric structure of the Universe over
large distances. The cases where some progress has been made with the use
of $d_A(z)$ include (1) the use of baryon acoustic oscillations seen in large-scale 
structure \cite{MeliaLopez:2017,Alcock:1979}; (2) the Sachs-Wolfe induced $\sim 10^\circ$
fluctuations seen in the cosmic microwave background \cite{Bennett:2003,Spergel:2003,MeliaLopez:2018};
(3) strong lensing systems, with and without time delays 
\cite{Refsdal:1964,Treu:2006,Biesiada:2010,Wei:2014a,MeliaWei:2015a};
and (4) galaxy clusters \cite{Sasaki:1996,Pen:1997,Melia:2016a}.
But several developments in our understanding of the compact structure 
in radio quasars have presented us with what appears to be a more reliable measuring 
rod, whose negligible evolution in the redshift range $0\lesssim z\lesssim 3$ permits 
precision cosmological testing over a larger fraction of the Universe's age than is 
feasible with these other methods, or even through the measurement of
the luminosity distance using Type Ia SNe. As we shall see below, these developments
follow primarily from our improved understanding of synchrotron self-absorption
processes near the central engine of active galactic nuclei (see, e.g., 
refs.~\cite{Blandford:1979,Melia:2009} and references cited therein) and the
identification of critical constraints on the observed characteristics of these
sources---principally their spectral index and luminosity---that permit the
selection of an appropriate sample with a more or less fixed size of the
emission region for cosmological testing 
\cite{Gurvits:1999,Cao:2015,Cao:2017a,Cao:2017b}.

Measuring the geometry of the Universe with standard rulers was first proposed
by Hoyle \cite{Hoyle:1959} over half a century ago, though it took several decades
before attempts were made to actually implement this proposal using actual sources.
And the earliest tests of cosmological models based on the observed redshift dependence
of the angular size of kpc-scale radio sources and galaxies were not successful
due to the lack of a reliable, well-defined standard ruler
\cite{Sandage:1988,Kapahi:1987,Barthel:1988,Neeser:1995,Singal:1993,Nilsson:1993}.
Some motivation to continue pursuing this quest finally came with a study of
double-lobed quasars within the redshift range $1.0\lesssim z\lesssim 2.7$,
which showed no change in apparent angular size with angular-diameter distance,
somewhat consistent with Friedmann-Robertson-Walker cosmology without any
significant evolution (see fig.~1 below) \cite{Buchalter:1998}.

It eventually became apparent that ultracompact radio sources are
more likely to produce standard measuring rods than the large-scale jets
in quasars and radio galaxies. The emission from these compact regions is
dominated by self-absorbed synchrotron emission \cite{Blandford:1979},
forming at least partially opaque features with angular diameters in 
the milliarcsecond (mas) range, and linear sizes of the order of $10$ parsecs
\cite{Kellerman:1981,Melia:2009}. Their significant
advantage over larger structures, such as galaxies and kpc-scale jets,
is that these central cores are much smaller than their parent active
galactic nuclei (AGN), so their ambient physical environment should
be similar from source to source and be reasonably stable, unlike the
variations one expects in the intergalactic medium over large distances and
times \cite{Kellermann:1993,Jackson:2008}. The compact 
structures in these sources therefore evolve principally under the
influence of the central engine itself, which is typically characterized
by only a few physical parameters, such as the mass of the black hole
and its spin. Dynamical timescales in such environments are only 
tens of years, much shorter than the age of the Universe. The compact
structures in radio quasars should therefore be free of long-term evolutionary
effects \cite{Gurvits:1999}.

In one of the more significant studies involving the compact 
structure of radio quasars, Gurvits, Kellermann \& Frey \cite{Gurvits:1999} showed that
a large sample of images taken with Very Large Baseline Interferometry (VLBI)
may be used to establish some general constraints
on cosmological parameters. This work has formed the basis of many subsequent
investigations \cite{Lima:2002,Zhu:2002,Chen:2003},
 leading to a second significant advancement with the use of these sources that we
shall discuss shortly \cite{Cao:2017a}.

A persistent complication with compact radio jets has been that they are found in 
a mixed population of radio galaxies and AGNs---quasars, BL Lacs, OVVs, etc.---so systematic 
differences are not always easy to disentangle from true cosmological variations. Recent 
work by Cao et al. \cite{Cao:2015}, however, has included the analysis of different AGN
sub-samples based on different source optical counterparts and lying in different
redshift ranges, leading to the conclusion that radio galaxies and quasars need
to be handled with distinct strategies. This result is the basis for the current study
of such sources, focusing only on compact structures in radio quasars
\cite{Cao:2017a,Cao:2017b}. The net outcome of this effort has been a
significant reduction in scatter to produce a reliable sample of compact structures 
in radio quasars for use as standard rulers, allowing us to carry out the study
reported in this {\it Letter}, which compares models in ways not previously feasible with
other measures of cosmological distance.

\section{Data and Analysis}
Following the suggestion by Gurvits et al. \cite{Gurvits:1999}
and Vishwakarma \cite{Vishwakarma:2001} that the exclusion of sources with low
luminosities $L$ and extreme spectral indices, $\alpha$, might curb the dependence
of the core size on the source luminosity and redshift, several workers have
refined the process of compiling from the many hundreds of available
VLBI images a reduced sample with manageable scatter, free of evolutionary
effects. 

With very long baseline interferometry, the signal from a distant
radio source is received at multiple radio telescopes across Earth's surface,
whose registration of intensities may then be correlated taking into account
the slightly different arrival times at the various facilities. The net result is
a combined observation made by a telescope with a baseline equal to the
maximum separation of the radio antennae. For the compact structure in
radio quasars, the characteristic angular size inferred from VLBI is
defined as
\begin{equation}
\theta_{\rm core}\equiv {2\sqrt{-\ln\Gamma\ln 2}\over \pi B}\;,
\end{equation}
where $B$ is the interferometer baseline and $\Gamma$ is the
ratio of total flux density to the correlated flux density \cite{Preston:1985}.
The linear size of the core may then be written
\begin{equation}
\ell_{\rm core}=\theta_{\rm core}(z) \times d_A(z)\;,
\end{equation}
where $d_A(z)$ is the model-dependent angular-diameter distance.

It is now understood that the dispersion in linear size is greatly
mitigated \cite{Gurvits:1999,Cao:2017a} by retaining only those sources with
$-0.38<\alpha<0.18$. Additionally, Cao et al. \cite{Cao:2017a} have recently
pointed to a strong dependence of the core size ${\ell}_{\rm core}$ on
luminosity, not just at the low end, but at the high end as well. Using the
parametrization ${\ell}_{\rm core}={\ell}_0\,L^\beta(1+z)^n$, where
${\ell}_0$ is a scaling constant, they demonstrated
that only a sub-sample of intermediate-luminosity radio quasars
($10^{27}\,{\rm W/Hz}<L<10^{28}\,{\rm W/Hz}$) have a core size with
negligible dependence on $L$ and $z$. For these objects, $\beta\approx 
10^{-4}$ and $|n|\approx 10^{-3}$, yielding a compilation of 
compact structures in radio quasars with a rather 
robust standard linear size.

\begin{figure}
\onefigure[width=3.4in]{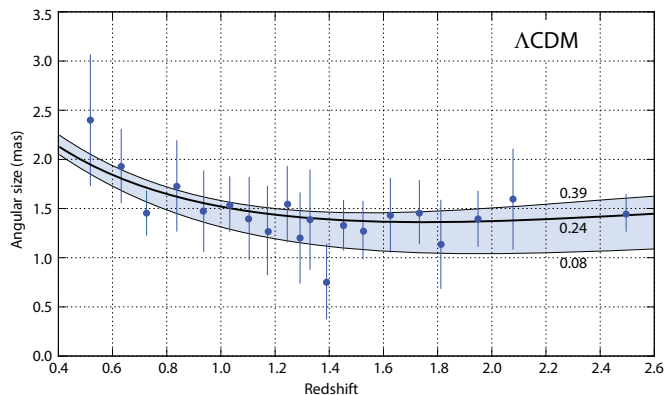}
\caption{Angular size of 140 individual compact structures in radio quasars 
binned into groups of 7, as a function of redshift. The data points represent the median
value in each bin. The thick solid curve is the optimized flat $\Lambda$CDM model,
with angular size constant $\eta=0.58\pm0.05$ and $\Omega_{\rm m}=0.24^{+0.1}_{-0.09}$
(see text). Also shown are the values of $\Omega_{\rm m}$ (i.e., $0.08$ and $0.39$)
setting the bounds of the $1\sigma$ shaded region (when $\eta$ is held constant).}
\label{fig.1}
\end{figure}

The data we use here were assembled by Jackson \& Jannetta \cite{JacksonJannetta:2006}
using the 2.29 GHz VLBI survey of Preston et al. \cite{Preston:1985} and additions
by Gurvits \cite{Gurvits:1994}, resulting in a catalog of 613 sources. In order for
us to extract the subsample with luminosities restricted to the range alluded to
above, we use the {\it Planck} optimized parameters \cite{Planck:2016} to estimate
the luminosity distance and thereby the value of $L$ from the measured total flux
density at $2.29$ GHz. These parameters are used merely to estimate $L$ for the
purpose of identifying the intermediate-luminosity sources, and are not otherwise
employed in the fitting procedure described below. We have carried out a
simple test to ensure that the model selection is not biased with this approach by
relaxing the constraint on $L$ by as much as $50\%$, which produced no discernable
effect. In addition, note that using the {\it Planck} parameters
would benefit $\Lambda$CDM, if at all, so an outcome favouring $R_{\rm h}=ct$
could not be viewed as having been facilitated by this approximation.
The sample is further reduced by
restricting $\alpha$ to the aforementioned range of spectral indices, producing a
final subsample of 140 sources for our study. These are binned into groups of 7 and,
following previously established convention \cite{Santos:2008}, we select the
median value to represent the angular size in each bin. These data are plotted
in figure~1, together with their $1\sigma$ errors estimated assuming a Gaussian
distributed variation in each bin.

Using the angular-diameter distance with these sources allows us to
study the geometry of the Universe in several unique ways. First, as long
as $\ell_{\rm core}$ is a true standard ruler, at least in an average
sense, we do not need to know its actual value for model selection because
we are merely sampling the ratio of scales at different redshifts. For the
same reason, we also do not need to know the Hubble constant, $H_0$,
further reducing the number of model parameters that need to be optimized.
And unlike Type Ia SNe, which may be detected only up to $z\sim 1.8$
\cite{Kowalski:2008}, compact radio jets have been mapped with 
VLBI as far out as $z\sim 4$ \cite{JacksonJannetta:2006,Preston:1985,Gurvits:1994}. 
In principle, we may therefore examine the geometry of the Universe over almost 
$80\%$ of its existence, compared to smaller fractions with other techniques. 
This follows from the fact that, in flat $\Lambda$CDM, a redshift
of $2.6$ corresponds to an age of about $2.5$ Gyr, which represents a look
back fraction $1-2.5/13.7\approx 0.82$, using today's age of $\sim 13.7$ Gyr.
Most critically, among the various distance measures in cosmology, $d_A(z)$
alone has a maximum value at a redshift $z_{\rm max}$ that varies from
one model to the next. The angular-diameter distance increases at first,
peaks at $z_{\rm max}$, and then decreases at higher redshifts, reaching
zero at the big bang. It is not difficult to understand why this
happens \cite{Melia:2013}. When we measure a lateral proper size, we
see the object as it was when it emitted the light reaching us today, and
since all sources were closer to us as we look back in time, their apparent
angular size $\theta_{\rm core}$ actually gets bigger as $z\rightarrow\infty$,
with the compensating effect that $d_A(z)$ ($\sim \theta_{\rm core}^{-1}$)
therefore gets smaller.

\begin{figure}
\onefigure[width=3.4in]{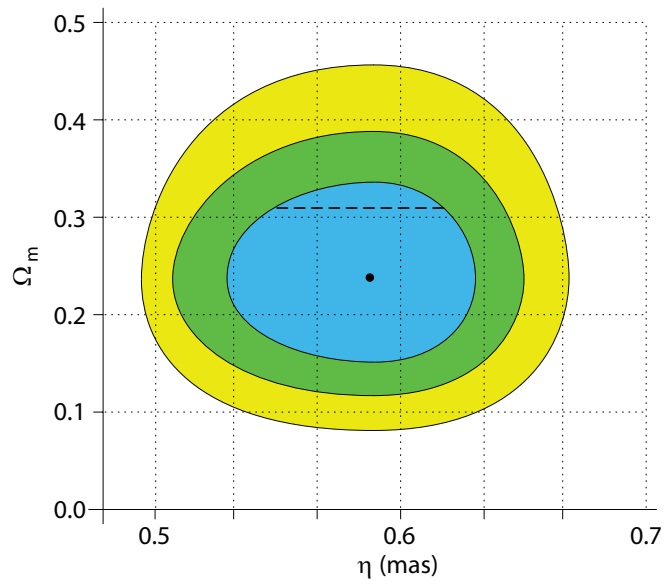}
\caption{One (blue), two (green), and three (yellow) $\sigma$ confidence regions
associated with the optimized parameters $\Omega_{\rm m}$ and $\eta$ in flat
$\Lambda$CDM (see Table~1). By comparison, the horizontal dashed line shows the
{\it Planck} \cite{Planck:2016} value $\Omega_{\rm m}=0.308$, which lies within
$1\sigma$ of the measurement based on compact structures in 
radio quasars.}
\label{fig.2}
\end{figure}

We will keep this analysis as simple and parameter-free as possible, concentrating
solely on what is absolutely needed in order for us to extract information concerning
the geometry of the Universe from the compact radio-jet data shown in figure~1. 
For a given core size $\ell_{\rm core}$, the predicted angular size of the
compact structure in radio quasars is obtained from Equation~(2), in terms of the
angular-diameter distance which, in flat $\Lambda$CDM, is given as
\begin{eqnarray}
d_A(z) &\hskip-0.1in=\hskip-0.1in& {c\over H_0}{1\over 1+z}\times\nonumber\\
&\null&\hskip-.8in\int_0^z {du\over\left[\Omega_{\rm m}(1+u)^3+
\Omega_{\rm r}(1+u)^4+\Omega_\Lambda(1+u)^{3+3w_\Lambda}\right]^{1/2}}\;.
\end{eqnarray}
In this expression, $\Omega_i$ is the energy density of species ``$i$" scaled to
today's critical density, $3c^2H_0^2/8\pi G$, and $w_\Lambda=-1$ is the
equation-of-state parameter for a cosmological constant. Given that radiation
energy density is relatively negligible up to $z\sim 3$, there are really
only two free parameters in the expression for $\theta_{\rm core}$ once we
marginalize over the unknowns $\ell_{\rm core}$ and $H_0$. We do this by
combining Equations~(2) and (3) and writing
\begin{equation}
\theta_{\rm core}(z) = \eta\,{1+z\over\mathcal{I}(z)}\;,
\end{equation}
where $\eta\equiv \ell_{\rm core}H_0/c$ and
\begin{equation}
\mathcal{I}(z)=\int_0^z {du\over\left[\Omega_{\rm m}(1+u)^3+1-\Omega_{\rm m}\right]^{1/2}}\;.
\end{equation}

Using standard $\chi^2$ minimization, we optimize the values of $\eta$ and
$\Omega_{\rm m}$ that produce the best fit, shown as a solid,
black curve in figure~1. The results of this fitting are summarized in Table~1,
and the corresponding $1\sigma$, $2\sigma$ and $3\sigma$ confidence regions
are plotted on the $\eta-\Omega_{\rm m}$ plane in figure~2. To further emphasize
the quality of the fit, we also show in figure~1 the $1\sigma$ confidence region
estimated by keeping $\eta$ fixed and allowing $\Omega_{\rm m}$ to vary. The
curves bounding this confidence region correspond to $\Omega_{\rm m}=0.08$
and $0.39$. We therefore find that the value of $\Omega_{\rm m}$ optimized
with fits to the compact radio-jet data is fully consistent with {\it Planck}
$\Lambda$CDM at a level of confidence of better than $1\sigma$.

\section{Discussion}
Let us now turn our attention to the primary goal
of using these sources for model selection purposes based solely on the inferred
geometry of the Universe. The turning point in the fitted $\theta_{\rm core}(z)$
function is a byproduct of $d_A(z)$'s unique maximum value at a
model-dependent redshift $z_{\rm max}$. The best-fit curve in figure~1 has a
turning point at $z_{\rm max}=1.69$, with a possible variation between 1.42
and 2.20 across the $1\sigma$ confidence region, a swing of about $46\%$ from
top to bottom.

We wish to understand what is so special about the values $\Omega_{\rm m}=0.24$
and $z_{\rm max}=1.69$ that the Universe would have `chosen' these to characterize
its geometry and expansion. Such questions have been asked repeatedly over the past
several decades as the analysis of each new dataset has unfolded. What has
emerged is an indication that $\Lambda$CDM may be lacking an additional constraint
on its constituents that may resolve the origin of these parameter values.

$\Lambda$CDM adopts the equation of state $p=w\rho$, with
$p=p_{\rm m}+ p_{\rm r}+p_{\rm de}$ and $\rho=\rho_{\rm m}+\rho_{\rm r}+
\rho_{\rm de}$, where $p_i$ and $\rho_i$ are the pressure and energy density,
respectively, of species ``$i$" in the cosmic fluid, with ``m" representing
(luminous and dark) matter, ``r" radiation and ``de" dark energy. The latter
may or may not be a cosmological constant. The expansion dynamics is based
on the assumption that $p_{\rm m}=0$, $p_{\rm r}=\rho_{\rm r}/3$ and
(typically) $p_{\rm de}=-\rho_{\rm de}$, while the densities
$\rho_{\rm m}\sim (1+z)^{-3}$, $\rho_{\rm r}\sim (1+z)^{-4}$ and
$\rho_{\rm de}\sim (1+z)^0$ each evolve with redshift independently of the
others. But the data appear to be telling us something different,
pointing to a coupling of the densities in order to preserve a constant
equation-of-state with $w=-1/3$, known as the `zero active mass' condition
in general relativity. In test after test, the predictions of $\Lambda$CDM
with this additional constraint, a model referred to as the $R_{\rm h}=ct$
Universe \cite{Melia:2007,MeliaAbdelqader:2009,MeliaShevchuk:2012,Melia:2016,Melia:2017}
in the literature, have been a better match to the data than those of
basic $\Lambda$CDM without it. These comparisons have been carried out
using a broad range of observations, from the angular correlation function
of the cosmic microwave background \cite{Melia:2014} and
high-$z$ quasars \cite{Melia:2013b,MeliaMcClintock:2015} in the early Universe, to
gamma ray bursts \cite{Wei:2014} and cosmic chronometers \cite{MeliaMaier:2013}
at intermediate redshifts, and to the relatively nearby Type Ia SNe \cite{Wei:2015}.
A recently compiled list of these comparative studies may be found in
Table~1 of \cite{Melia:2017b}.

\begin{figure}
\onefigure[width=3.4in]{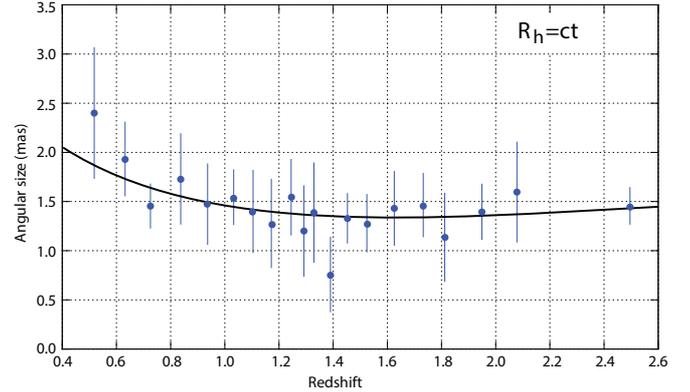}
\caption{Same as figure~1, except now comparing the data with the curve predicted
by the $R_{\rm  h}=ct$ Universe. Once the Hubble constant $H_0$ and the size
of the compact structure are marginalized, there are no free parameters
left with which to optimize the fit. Even without this flexibility, however,
this curve accounts for the data better than $\Lambda$CDM (see Table~1).}
\label{fig.3}
\end{figure}

The angular-size data presented in figure~1 allows us examine this growing
body of evidence from an entirely different perspective, because we have
never had an opportunity before the identification of this sample of compact radio
structures by key workers in this field, including Gurvits \cite{Gurvits:1994},
Jackson \cite{Jackson:2008}, Cao et al. \cite{Cao:2017a,Cao:2017b}
and others, of testing the prediction that $d_A(z)$
ought to have a maximum at a finite redshift $z_{\rm max}$. When we
impose the additional constraint $w=-1/3$ on $\Lambda$CDM, Equation~(5)
is simply
\begin{equation}
\mathcal{I}(z) = \ln(1+z)\;.
\end{equation}
The angular size of the compact radio jets predicted by this model is
now shown together with the data in figure~3. It is virtually indistinguishable
from the best-fit $\Lambda$CDM curve in figure~1. Indeed, the reduced $\chi^2_{\rm dof}$'s
for these two fits are identical (see Table~1). But what is particularly
significant is the fact that the theoretical curve in figure~3 has no free
parameters (other than the `nuisance' variable $\eta$, which is common to
both models). In model selection, one must take the different number of
free parameters into account when evaluating relative likelihoods. In
cosmology, this is now routinely done using information criteria, such as
the Bayes Information Criterion (BIC) applicable to large samples
\cite{MeliaMaier:2013}, defined as ${\rm BIC}=\chi^2+k\ln n$, where $k$
is the number of free parameters and $n$ is the sample size. In a head
to head comparison between two models, the BIC yields the relative
probability of either being `closer' to the truth and, as one can see
from the outcomes listed in Table~1, the use of $d_A(z)$ with the compact
radio-jet data favours $R_{\rm h}=ct$ over $\Lambda$CDM.

\begin{table}
\caption{Model Selection Using the Compact Structure of Radio Quasars}
\footnotesize
\begin{center}
\begin{tabular}{lccccc}
\label{Tab:fits}
Model & $\Omega_{\rm m}$ & $\eta$ & $\chi^2_{\rm dof}$ & BIC & Probability \\
\hline\hline
&&&&& \\
$\Lambda$CDM & $0.24^{+0.1}_{-0.09}$ & $0.58\pm 0.05$  & 0.31 & 11.6 & $19.8\%$ \\
&&&&& \\
$R_h=ct$ & --- & $0.5^{+0.03}_{-0.02}$  & 0.31 & 8.8 & $80.2\%$ \\
&&&&& \\
\hline\hline
\end{tabular}
\end{center}
\end{table}

We are therefore seeing a strong confirmation of previous results
based on other kinds of measurement. This outcome suggests that the
optimized value of $\Omega_{\rm m}$ in $\Lambda$CDM arises because
the formulation of $w$ in this model needs it to mimic the integral
in Equation~(6) associated with the zero active mass condition $w=-1/3$.
But the most convincing evidence in support of this conclusion comes
from an evaluation of $z_{\rm max}$ in $R_{\rm h}=ct$, a new probe
of the Universe's geometry---never seen before the recent 
work of Cao et al. \cite{Cao:2017a,Cao:2017b} with any other kind of
measurement. From Equations~(4) and (6) one finds that $z_{\rm max}=1.718$
when the zero active mass condition is imposed on $\Lambda$CDM.
This value lies within $1.6\%$ of the turning point in $\theta_{\rm core}$
found with the formulation of $w$ in the standard model suggesting, once
again, that the measured value of $z_{\rm max}$ is not random at all, but
is a direct consequence of the Universe's expansion at a rate consistent
with the zero active mass condition.

Finally, let us consider what the optimized value of $\eta$
in Table~1 implies for the Hubble constant $H_0$, should the core size
${\ell}_{\rm core}$ be known from other data. A difficulty generally arises 
in the optimization of $H_0$ due to its model dependence. There is no 
universal way of measuring its value without assuming a particular model.
The Hubble constant has been measured locally, e.g., using Cepheid
variables, though it disagrees with the {\it Planck} value by more
than $9\%$.  It is not yet clear why this happens, but some have
speculated that a local ``Hubble bubble" (Shi 1997; 
Keenan et al. 2013; Romano 2017) may be influencing the local dynamics 
within a distance $\sim 300$ Mpc (i.e., $z\lesssim 0.07$). If true, such a 
fluctuation might lead to anomalous velocities within this region, 
causing the nearby expansion to deviate somewhat from 
a pure Hubble flow. For consistency, $H_0$ must therefore
be measured on large, smoothed scales. 

With our value of $\eta$, $H_0$ may be inferred once ${\ell}_{\rm core}$
is known. An estimate of its value was made recently by ref.~\cite{Cao:2017b},
who used measurements of the expansion rate $H(z)$ based on cosmic chronometers
to break the degeneracy between ${\ell}_{\rm core}$ and the Hubble constant. 
Their analysis estimated the core size to be ${\ell}_{\rm core}\sim 11.03\pm0.25$ pc.
Thus, for the optimized value $\eta=0.5^{+0.03}_{-0.02}$ in $R_{\rm h}=ct$ (see Table~1), 
the implied Hubble constant is $H_0=66.0^{+4.0}_{-2.6}$ km s$^{-1}$ Mpc$^{-1}$. As of today, $H_0$ in 
$R_{\rm h}=ct$ has been measured 4 times: $63.2 \pm 1.6$ km s$^{-1}$ Mpc$^{-1}$ \cite{MeliaMaier:2013},
$63.3 \pm 7.7$ km s$^{-1}$ Mpc$^{-1}$ \cite{MeliaMcClintock:2015}, $62.3\pm 1.5$ km s$^{-1}$ Mpc$^{-1}$ 
\cite{Wei:2017} and $63.0 \pm 2.0$ km s$^{-1}$ Mpc$^{-1}$ \cite{MeliaYenna:2018}.
Cao et al.'s estimate of ${\ell}_{\rm core}$ therefore yields a Hubble constant $H_0$ for 
$R_{\rm h}=ct$ consistent with these previous measurements. 

The corresponding Hubble constant in $\Lambda$CDM is $76.6\pm6.6$ km s$^{-1}$ Mpc$^{-1}$.
But while this value is still consistent with the Planck optimization, it is nonetheless somewhat 
on the high side, in slight tension with the estimates made in ref.~\cite{Cao:2017b}. It is
very instructive to examine the cause of this non-trivial difference.
There are actually two principal reasons why our inferred value of $H_0$ for
$\Lambda$CDM does not agree completely with that obtained earlier by
ref.~\cite{Cao:2017b}. 

First, while Cao et al. carried out a $\chi^2$-minimization 
using the individual quasars in our sample of 120, our approach
calls for the binning of these sources into 20 redshift intervals before
optimizing the model parameters. A quick inspection of their data
and best fit curves shows that many of the individual sources lie 
several $\sigma$'s away from the theoretical curves. In 
other words, the reported errors are far too small to represent the 
actual scatter in the data. For this reason, we have chosen to bin
the individual sources and use population variance based on
assumed Gaussian variation within each bin to more reliably
estimate the error associated with each data point. Not 
surprisingly, our errors are larger than those reported for 
each individual source because they better reflect
the overall scatter in the data. A quick inspection of figs.~1 and 3 
shows that all but one of the data in these plots lie within $1\sigma$ 
of the best-fit curves. We suggest that carrying out a model optimization 
with these binned data is therefore more reliable than simply 
trying to do this with data whose errors are unrealistically 
small. 

Second, and more importantly, Cao et al. \cite{Cao:2017b} optimized 
${\ell}_{\rm core}$ and $H_0$ separately, while (as noted above), we 
optimize the sole parameter $\eta$. Why is this important? Cao et al. 
did not base their parameter optimization solely on the quasar-core
data. Since they needed additional information to separate 
these two unknowns, they combined their angular-size measurements
with observations of $H(z)$ based on cosmic chronometers, as noted 
earlier. As such, their optimized parameters reflect the joint
analysis of several different data sets, as opposed to just the
quasar-core observations---the principal focus in this paper. 
Each approach has its advantages, of course. Ours allows 
model selection to be carried out based solely on the quasar-core 
data. This is not a trivial step, because each kind of measurement 
should be studied on its own, not only in joint analyses with
other observations that may introduce unknown biases.

\section{Conclusion}
Needless to say, the identification of the compact structure in 
radio quasars as standard
rulers has opened up an entirely new chapter in cosmology. With them, we
may now map the geometry of the Universe well beyond the reach of Type
Ia SNe, sampling even the epoch during which the apparent size of sources
increases with redshift, an effect not seen with any other kind of
measurement probe. The results thus far point to the zero active mass
condition in general relativity as the influence guiding the Universe's
expansion. Developing this notion further, and testing it with even higher
precision measurements, promises a very exciting future in observational
cosmology.

\acknowledgments
I am very grateful to the anonymous referee, who has provided
many suggestions for improving the presentation in this paper. I am also 
grateful to Amherst College for its support through
a John Woodruff Simpson Lectureship, and to Purple Mountain Observatory in Nanjing,
China, for its hospitality while part of this work was being carried out. This work
was partially supported by grant 2012T1J0011 from The Chinese Academy of Sciences
Visiting Professorships for Senior International Scientists, and grant GDJ20120491013
from the Chinese State Administration of Foreign Experts Affairs.

\end{document}